\documentclass{emulateapj}
\slugcomment{Submitted September 8, 2012; Accepted January 9, 2013}

\shorttitle{Nuclear Radio Jet of NGC 4258}
\shortauthors{Doi et al.}

\begin{document}

\title{Nuclear Radio Jet from a Low-Luminosity Active Galactic Nucleus in NGC 4258}

\author{Akihiro Doi\altaffilmark{1,2}, Kotaro Kohno\altaffilmark{3,4}, Kouichiro Nakanishi\altaffilmark{5,6,7}, Seiji Kameno\altaffilmark{8}, Makoto Inoue\altaffilmark{9}, Kazuhiro Hada\altaffilmark{10},\\ and Kazuo Sorai\altaffilmark{11}}

\altaffiltext{1}{The Institute of Space and Astronautical Science, Japan Aerospace Exploration Agency, 3-1-1 Yoshinodai, Chuou-ku, Sagamihara, Kanagawa 252-5210, Japan}
\altaffiltext{2}{Department of Space and Astronautical Science, The Graduate University for Advanced Studies (SOKENDAI), 3-1-1 Yoshinodai, Chuou-ku, Sagamihara, Kanagawa 252-5210, Japan}
\altaffiltext{3}{Institute of Astronomy, The University of Tokyo, 2-21-1 Osawa, Mitaka, Tokyo 181-0015, Japan}
\altaffiltext{4}{Research Center for the Early Universe, School of Science, University of Tokyo, 7-3-1 Hongo, Bunkyo, Tokyo 113-0033, Japan}
\altaffiltext{5}{National Astronomical Observatory of Japan, 2-21-1 Osawa, Mitaka, Tokyo 181-8588, Japan}
\altaffiltext{6}{Joint ALMA Observatory, Alonso de Cordova 3107, Vitacura, Santiago, 763-0355, Chile}
\altaffiltext{7}{Department of Astronomical Science, The Graduate University for Advanced Studies (SOKENDAI), 2-21-1 Osawa, Mitaka, Tokyo 181-0015, Japan}
\altaffiltext{8}{Department of Physics, Faculty of Science, Kagoshima University, 1-21-35 Korimoto, Kagoshima, Kagoshima 890-0065, Japan}
\altaffiltext{9}{Academia Sinica Institute of Astronomy and Astrophysics, P.O. Box 23-141, Taipei 10617, Taiwan}
\altaffiltext{10}{INAF, Istituto di Radioastronomia, via Gobetti 101, Bologna 40129, Italy}
\altaffiltext{11}{Department of Physics, Graduate School of Science, Hokkaido University, Kita 10 Nishi 8, Sapporo 060-0810, Japan}

\begin{abstract}
The nearby low-luminosity active galactic nucleus~(LLAGN) NGC~4258 has a weak radio continuum component at the galactic center.  
We investigate its radio spectral properties on the basis of our new observations using the Nobeyama Millimeter Array at 100~GHz and archival data from the Very Large Array~(VLA) at 1.7--43~GHz and the James Clerk Maxwell telescope at 347~GHz.  
The NGC~4258 nuclear component exhibits (1)~an intra-month variable and complicated spectral feature at 5--22~GHz and (2)~a slightly inverted spectrum at 5--100~GHz ($\alpha \sim 0.3$; $F_\nu \propto \nu^{\alpha}$) in time-averaged flux densities, which are also apparent in the closest LLAGN M81.   
These similarities between NGC~4258 and M81 in radio spectral natures in addition to previously known core shift in their AU-scale jet structures produce evidence that the same mechanism drives their nuclei.   
We interpret the observed spectral property as the superposition of emission spectra originating at different locations with frequency-dependent opacity along the nuclear jet.  
Quantitative differences between NGC~4258 and M81 in terms of jet/counter jet ratio, radio loudness, and degree of core shift can be consistently understood by fairly relativistic speeds ($\Gamma \ga 3$) of jet and their quite different inclinations.  
The picture established from the two closest LLAGNs is useful for understanding the physical origin of unresolved and flat/inverted spectrum radio cores that are prevalently found in LLAGNs, including Sgr~A*, with starved supermassive black holes in the present-day universe.   
\end{abstract}


\section{INTRODUCTION}\label{section:introduction} 
Most of the supermassive black hole activity in the present-day universe is primarily in the lower end of the luminosity function of active galactic nuclei~(AGNs) in the form of low-luminosity AGNs \citep[LLAGNs; for reviews, see][]{Ho:1997,Ho:2008} with bolometric luminosities $L_\mathrm{bol} \la 10^{42}$~erg~s$^{-1}$ down to nearly quiescent systems such as Sgr~A* at our Galactic center.     
Most LLAGNs are extremely sub-Eddington systems ($L_\mathrm{bol}/L_\mathrm{Edd}<10^{-3}$, where $L_\mathrm{Edd}$ is the Eddington luminosity).  
The low luminosities and spectral energy distributions~(SEDs) of LLAGNs \citep{Ho:1999a,Eracleous:2010,Younes:2012} have been modeled by the combination of 
(i)~an inner advection-dominated accretion flow~\citep[ADAF;][]{Narayan:1994}, which is a radiatively inefficient, optically thin, and geometrically thick accretion flow at low accretion rates; 
(ii) an outer truncated disk \citep{Quataert:1999} of a standard optically thick, geometrically thin accretion disk \citep{Shakura:1973}; and 
(iii) a jet \citep{Falcke:1995,Yuan:2005a}, 
as proposed by several studies \citep[e.g.,][and references therein]{Nemmen:2006,Nemmen:2011,Yu:2011}.

The ADAF model predicts an SED including a ``sub-millimeter bump'' that results from the synchrotron emission of thermal hot ($\sim10^9$--$10^{10}$~K) electrons.  A highly inverted spectrum \citep[$\alpha=0.4$;][in $S_\nu \propto \nu^\alpha$, where $\alpha$ is the spectral index and $S_\nu$ is the flux density at frequency $\nu$]{Mahadevan:1997} is expected in lower-frequency radio bands.    
However, observed spectra in LLAGNs tend to be flat or slightly inverted ($\alpha \approx -0.2$ to $+0.2$) in the central components \citep{Nagar:2001,Doi:2005,Doi:2011} as well as the spectrum of Sgr~A* at $\la10$~GHz \citep{Falcke:1998,An:2005}.  
Furthermore, only the thermal process in the ADAF model cannot provide observed radio luminosities adequately in many cases \citep{Fabbiano:2003,Doi:2005a,Wu:2005,Wu:2007}.   
Thus, these are the reason why the jet component is needed in SED modeling for LLAGNs.  
It is also argued that the entire SED of LLAGNs is practically jet-dominated by synchrotron emission and up-scattered inverse Compton component rather than by the accretion flow or disk \citep[in particular at very low luminosities $L_\mathrm{X}/L_\mathrm{Edd} \la 10^{-6}$;][]{Yuan:2005}, according to studies for individual sources (NGC~4258; \citealt{Yuan:2002}, M81; \citealt{Markoff:2008}) and many samples in the statistical sense \citep{Merloni:2003,Falcke:2004,Kording:2006,Yuan:2009,de-Gasperin:2011,Plotkin:2012}.  
Thus, the jet component is presumably essential in the energetics at the low state of AGNs.

Very-long-baseline interferometry~(VLBI) images at milli-arcsecond~(mas) resolutions have sometimes revealed elongated radio structures similar to sub-parsec scale jets in LLAGNs \citep{Falcke:2000,Nagar:2005}.  
With the exception of a handful of well-known nearby radio galaxies and Seyfert galaxies with jets extending to kpc scales \citep{Wrobel:1991,Ho:2001}, 
most of the power is concentrated in an unresolved core of $\la10^3$Rs--$10^4$Rs (Rs is the Schwarzschild radius) with high brightness temperatures of $\la10^{6}$--$10^{11}$~K \citep{Ulvestad:2001,Anderson:2004,Filho:2004,Krips:2007}. 
Radio flux variability observed in several LLAGNs in total flux on typical timescales of a few days also indicates $\la1000$Rs in size \citep{Anderson:2005}.  
Thus, the radio emitting origin has been still unknown for most LLAGNs from an observational point of view.   
One of the rare cases in which a nuclear radio component has been spatially resolved is M81 \citep{Bietenholz:1996}, which is the nearest (3.6~Mpc; \citealt{Freedman:2001}) LLAGN in the VLBI-observed sample of \citet{Nagar:2005}.    
M81 contains a type 1.5 Seyfert nucleus \citep{Ho:1997} with a 2--10~keV X-ray luminosity of $\sim 1.5$--$4 \times 10^{40}$~erg~s$^{-1}$ \citep{Ishisaki:1996,Markoff:2008} and a black hole with a mass of $7\times10^7\ M_{\sun}$ \citep{Devereux:2003}.
The radio nucleus consists of a bright core plus weak one-sided jet-like elongation with a scale of $\sim1$~mas ($3 \times 10^3$Rs) at 1--43~GHz in VLBI images \citep{Bietenholz:2004,Ros:2012}.     
Observed radio properties are characterized by as follows.  
(1)~Core shift: the position of radio brightness peak appears shifted depending on observing frequency at 1.7--8.4~GHz; the line of sight toward the putative black hole is opaque at these radio frequencies \citep{Bietenholz:2004,Marti-Vidal:2011}.  
Therefore, almost all radio fluxes should be attributed to AU-scale jets rather than an accretion flow.  
(2)~Flux variability: M81 is the most well-studied LLAGN in total flux as well; a bright radio nucleus ($\sim70$--400~mJy) exhibits rapid and large amplitude intraday fluctuation \citep{Ho:1999,Sakamoto:2001} and on the timescale of several weeks and years \citep{Bietenholz:2000,Markoff:2008,Marti-Vidal:2011}.  
(3)~Inverted spectrum: a centimeter-to-submillimeter (cm-to-submm) spectrum is slightly inverted on time average \citep[$\alpha \approx 0.3$;][]{Reuter:1996,Doi:2011} with a possible turnover frequency at 100--230~GHz \citep{Schodel:2007}.   
(4)~Spectral variability: a complexly variable spectral profile was observed at meter to submm wavelengths (235~MHz--345~GHz) in quasi-simultaneous observations over six months \citep{Markoff:2008}.  
Thus, the radio properties (2)--(4) at arcsec resolutions are ascribable to nuclear jets whose structure is spatially resolved as (1).   
For this reason, M81 may be a Rosetta Stone for understanding the origin of unresolved radio cores in other LLAGNs and Sgr~A*.

NGC~4258 contains a type~1.9 Seyfert nucleus \citep{Ho:1997} with a 2--10~keV X-ray luminosity of (4.2--17.4)~$\times10^{40}$~erg~s$^{-1}$ \citep{Makishima:1994,Fruscione:2005} and a black hole with a mass of $3.9\times10^7\ M_{\sun}$ \citep{Herrnstein:1999}.  
NGC~4258 is the second nearest \citep[7.2~Mpc;][]{Herrnstein:1999} LLAGN in the VLBI-observed sample of \citet{Nagar:2005}.  
Thereby, NGC~4258 in addition to M81 constitute rare cases in which a nuclear radio structure is spatially resolved \citep{Miyoshi:1995}.  
VLBI imaging at $\sim1.5$~GHz showed a two-sided nuclear jet extending in the north--south direction \citep{Cecil:2000} that appears physically connected with the kpc-scale jet-like radio structures (``anomalous arms''; e.g., \citealt{van-der-Kruit:1972}). 
The nuclear radio structure has been thoroughly investigated at 22~GHz through VLBI observations of water masers \citep{Herrnstein:2005,Argon:2007}.  
A two-sided nuclear jet with a scale of $\sim1$~mas ($9\times10^3$Rs) 
upwells from the dynamical center of a putative disk of Keplerian rotating water masers in a disk with a nearly edge-on orientation.   
VLBI images provided a $3\sigma$ upper limit of $220\ \mu$Jy on 22~GHz at the dynamical center; 
most of the nuclear radio continuum flux comes from the jet, which is located significantly off the dynamical center \citep{Herrnstein:1998}.  
The jet exhibits a flux variation of $\sim100$\% on the timescales of a few weeks or longer (\citealt{Herrnstein:1997}, see also Appendix~\ref{section:previous_studies}).  
Thus, the observed properties in terms of innermost jet structure and flux variability of the nuclear component are very similar to those of M81 ((1) and (2)).   
However, the radio spectral properties are known in much less detail (Appendix~\ref{section:previous_studies}): (3)~a slightly inverted spectrum?  (4)~a variable spectral profile?     
To establish the archetype of LLAGNs by using M81 and NGC~4258, the radio spectral properties of NGC~4258 nucleus should be investigated and compared with those of M81 at cm-to-submm wavelengths.  

In this paper, we present the radio continuum spectrum and its variability for the NGC~4258 nucleus in the cm-to-submm bands at arcsecond resolutions.  This study also includes imaging of extended components, such as extended synchrotron jets and dust emission in a host galaxy, for the purpose of estimation of contamination.  The observations and data reductions are described in Section~\ref{section:observationsanddatareductions}.  The results are presented in Section~\ref{section:results}.  The physical origins of observed radio properties and comparisons with M81 are discussed in Section~\ref{section:discussion}.  Finally, we summarize in Section~\ref{section:summary}.  For the distance of 7.2~Mpc to NGC~4258, $1\arcsec$ corresponds to 35~pc.

\section{Observations and data reductions}\label{section:observationsanddatareductions}
For the purpose of revealing radio spectral variability and its average nature for this variable radio nucleus in NGC~4258, we obtained multi-epoch data.  
We carried out new observations of 15 epochs using the Nobeyama Millimeter Array~(NMA) at the Nobeyama Radio Observatory~(NRO) at $\sim100$~GHz (Section~\ref{section:NMAobservation}).  We retrieved the 1.7--43~GHz archival data of the Very Large Array~(VLA) at the National Radio Astronomy Observatory~(NRAO) for 13 epochs in total (Section~\ref{section:VLAdata}).  The list of these observations is presented in Table~\ref{table:obslist}.  
We reduced the 347 GHz archival data of the James Clerk Maxwell telescope (JCMT) on Mauna Kea (Section~\ref{section:JCMTdata}), which was, as a result, devoted to estimation of dust contamination in photometry toward the active nucleus because the nuclear region was dominated by dust emission at this frequency (Section~\ref{section:result_submm}).

\begin{table*}
\caption{Data List of VLA and NMA Observations}\label{table:obslist}
{\scriptsize  
\begin{center}
\begin{tabular}{lcccccrc} 
\tableline\tableline
Date & Project & Array & $\nu$ & $\theta_\mathrm{maj}$ & $\theta_\mathrm{min}$ & P.A. & $\sigma$ \\
 & code & config. & (GHz) & (\arcsec) & (\arcsec) & (\degr) & (mJy beam$^{-1}$) \\
(1) & (2) & (3) & (4) & (5) & (6) & (7) & (8) \\
\tableline
07 Dec 1996 & AH594A & VLA-A & 22.460 & 0.10 & 0.09 & 34 & 0.12  \\
 &  &  & 14.940 & 0.15 & 0.14 & $-8$ & 0.10  \\
 &  &  & 8.460 & 0.29 & 0.23 & $-12$ & 0.04  \\
 &  &  & 4.860 & 0.47 & 0.43 & $-18$ & 0.06  \\
22 Dec 1996 & AH594B & VLA-A & 22.460 & 0.11 & 0.09 & 28 & 0.11  \\
 &  &  & 14.940 & 0.15 & 0.13 & $-7$ & 0.11  \\
 &  &  & 8.460 & 0.28 & 0.23 & $-14$ & 0.05  \\
 &  &  & 4.860 & 0.47 & 0.43 & $-21$ & 0.06  \\
29 Dec 1996 & AH594C & VLA-A & 22.460 & 0.10 & 0.09 & 22 & 0.13  \\
 &  &  & 14.940 & 0.15 & 0.14 & $-2$ & 0.11  \\
 &  &  & 8.460 & 0.28 & 0.24 & $-11$ & 0.06  \\
 &  &  & 4.860 & 0.46 & 0.44 & $-10$ & 0.07  \\
31 Dec 1996 & AH594D & VLA-A & 22.460 & 0.10 & 0.09 & 29 & 0.17  \\
 &  &  & 14.940 & 0.15 & 0.14 & $-6$ & 0.15  \\
 &  &  & 8.460 & 0.24 & 0.20 & $-15$ & 0.05  \\
 &  &  & 4.860 & 0.47 & 0.43 & $-19$ & 0.06  \\
04 Jan 1997 & BR0043 & VLA-A & 15.365 & 0.16 & 0.14 & $-87$ & 0.11  \\
 &  &  & 8.415 & 0.70 & 0.27 & $-66$ & 0.07  \\
 &  &  & 4.985 & 0.92 & 0.44 & $-69$ & 0.07  \\
05 Jan 1997 & AH594E & VLA-A & 22.460 & 0.10 & 0.09 & 28 & 0.18  \\
 &  &  & 14.940 & 0.16 & 0.14 & $-6$ & 0.15  \\
 &  &  & 8.460 & 0.29 & 0.23 & $-13$ & 0.05  \\
 &  &  & 4.860 & 0.47 & 0.43 & $-20$ & 0.07  \\
07 Jan 1997 & BG0062 & VLA-A & 1.665 & 1.23 & 1.41 & 83 & 0.04  \\
 &  &  & 22.460 & 0.17 & 0.11 & $-69$ & 0.23  \\
06 Mar 1997 & BM0056\tablenotemark{a} & VLA-B & 22.265 & 0.35 & 0.31 & $-66$ & 0.10  \\
 &  &  & 8.460 & 1.88 & 0.83 & 75 & 0.06  \\
 &  &  & 4.860 & 3.54 & 1.43 & 69 & 0.12  \\
19 Mar 1997 & AH594F & VLA-B & 22.460 & 0.43 & 0.32 & 82 & 0.20  \\
 &  &  & 8.460 & 1.42 & 0.90 & 85 & 0.12  \\
 &  &  & 4.860 & 2.24 & 1.47 & $-83$ & 0.14  \\
28 Feb 1998 & AG0527 & VLA-A & 43.340 & 0.05 & 0.04 & $-13$ & 0.41  \\
 &  &  & 22.460 & 0.11 & 0.09 & 86 & 0.33 \\
05 Sep 1998 & BM0112\tablenotemark{a} & VLA-B & 22.265 & 0.40 & 0.33 & 87 & 0.16  \\
 &  &  & 14.940 & 1.97 & 0.45 & 56 & 0.16  \\
 &  &  & 8.460 & 2.41 & 0.81 & 62 & 0.07  \\
 &  &  & 4.860 & 3.74 & 1.41 & 64 & 0.10  \\
21 Dec 2000 & AN0097 & VLA-A & 43.340 & 0.09 & 0.04 & 76 & 0.51  \\
 &  &  & 8.460 & 0.42 & 0.24 & 79 & 0.11  \\
21 Dec 2001 &  & NMA-D & 100.777 & 7.5 & 6.2 & $-72$ & 0.89  \\
21 Mar 2002 &  & NMA-C & 100.777 & 8.8 & 4.7 & $-36$ & 3.00  \\
22 Dec 2003 &  & NMA-C & 100.777 & 4.3 & 3.5 & $-2$ & 0.94  \\
14 Jan 2004 &  & NMA-AB & 100.777 & \ldots & \ldots & \ldots & 1.96  \\
31 Mar 2004 &  & NMA-D & 100.777 & 6.6 & 5.7 & $-76$ & 0.93  \\
01 Apr 2004 &  & NMA-D & 100.777 & \ldots & \ldots & \ldots & 2.76  \\
05 Apr 2005 &  & NMA-C & 100.777 & 3.8 & 3.0 & $-33$ & 0.84  \\
08 Apr 2005 &  & NMA-C & 100.777 & 4.4 & 3.6 & $-30$ & 1.49  \\
13 May 2005 &  & NMA-D & 95.729 & 7.5 & 6.2 & $-88$ & 0.87  \\
14 May 2005 &  & NMA-D & 95.729 & 9.5 & 7.9 & $-70$ & 1.93  \\
15 May 2005 &  & NMA-D & 95.729 & 9.9 & 6.8 & 76 & 0.94  \\
25 Mar 2006 &  & NMA-D & 100.777 & \ldots & \ldots & \ldots & 3.12  \\
29 Mar 2006 &  & NMA-D & 100.777 & \ldots & \ldots & \ldots & 3.12  \\
31 Mar 2006 &  & NMA-D & 100.777 & \ldots & \ldots & \ldots & 2.37  \\
01 Apr 2006 &  & NMA-D & 100.777 & \ldots & \ldots & \ldots & 2.86  \\
\tableline
\end{tabular}
\end{center}
\tablecomments{Column~1: observation date; Column~2: project code; Column~3: telescope and array configuration; Column~4: observing frequency; Columns~5--7: FWHMs in major and minor axes for synthesized beam and position angle of the major axis; Column~8: image rms noise.}
}

\tablenotetext{a}{VLA participated in VLBI observation as a phased-up element, and the VLBI results have been published \citep{Argon:2007}.}  
\end{table*}

\begin{table*}
\caption{Flux Densities of Radio Continuum Emission at the Nucleus of NGC 4258}\label{table:n4258result}
{\scriptsize  
\begin{center}
\begin{tabular}{lccccccccc} 
\tableline\tableline
Date & $S_\mathrm{1.7}$ & $S_\mathrm{5}$ & $S_\mathrm{8.4}$ & $S_\mathrm{15}$ & $S_\mathrm{22}$ & $S_\mathrm{43}$ & $S_\mathrm{100}$ & $S_\mathrm{347}$ & $\alpha^{22}_5$ \\
 & (mJy) & (mJy) & (mJy) & (mJy) & (mJy) & (mJy) & (mJy) & (mJy) \\
(1) & (2) & (3) & (4) & (5) & (6) & (7) & (8) & (9) & (10) \\
\tableline
07 Dec 1996 &  & $1.8 \pm 0.1$ & $1.9 \pm 0.1$ & $2.2 \pm 0.2$ & $2.4 \pm 0.3$ &  &  &  & $0.21 \pm 0.04$ \\
22 Dec 1996 &  & $2.0 \pm 0.1$ & $2.2 \pm 0.1$ & $2.2 \pm 0.2$ & $2.2 \pm 0.2$ &  &  &  & $0.06 \pm 0.06$ \\
29 Dec 1996 &  & $1.9 \pm 0.1$ & $2.3 \pm 0.1$ & $2.7 \pm 0.2$ & $2.5 \pm 0.3$ &  &  &  & $0.21 \pm 0.07$ \\
31 Dec 1996 &  & $1.9 \pm 0.1$ & $2.3 \pm 0.1$ & $2.7 \pm 0.3$ & $3.1 \pm 0.4$ &  &  &  & $0.32 \pm 0.01$ \\
04 Jan 1997 &  & $1.8 \pm 0.2$ & $2.0 \pm 0.1$ & $3.5 \pm 0.2$ &  &  &  &  & $0.64 \pm 0.23$ \\
05 Jan 1997 &  & $2.3 \pm 0.1$ & $2.3 \pm 0.1$ & $2.6 \pm 0.3$ & $3.6 \pm 0.4$ &  &  &  & $0.19 \pm 0.12$ \\
07 Jan 1997 & $2.0\pm0.1$\tablenotemark{b} &  &  &  & $3.7 \pm 0.4$ \\
06 Mar 1997 &  & $3.1 \pm 0.3$ & $2.7 \pm 0.2$ &  & $5.4 \pm 0.3$ &  &  &  & $0.62 \pm 0.19$ \\
19 Mar 1997 &  & $2.7 \pm 0.3$ & $2.8 \pm 0.3$ &  & $3.2 \pm 0.4$ &  &  &  & $0.10 \pm 0.01$ \\
28 Feb 1998 &  &  &  &  & $4.4 \pm 0.6$ & $7.1 \pm 0.9$ \\
18 Mar 1998 &  &  &  &  &  &  &  & $93.7 \pm 20.8$\tablenotemark{d} \\
05 Sep 1998 &  & $2.5 \pm 0.3$ & $2.2 \pm 0.1$ & $3.6 \pm 0.4$ & $3.0 \pm 0.4$ &  &  &  & $0.26 \pm 0.21$ \\
05 Sep 1999\tablenotemark{a} &  & $2.1 \pm 0.2$ & $1.8 \pm 0.1$ & $2.5 \pm 0.3$ &  &  &  &  & $0.19 \pm 0.24$ \\
21 Dec 2000 &  &  & $2.6 \pm 0.2$ &  &  & $6.8 \pm 1.0$ \\
21 Dec 2001 &  &  &  &  &  &  & $10.0 \pm 1.3$ \\
21 Mar 2002 &  &  &  &  &  &  & $7.4 \pm 1.6$ \\
22 Dec 2003 &  &  &  &  &  &  & $5.2 \pm 1.1$ \\
14 Jan 2004 &  &  &  &  &  &  & $<5.9$ \\
31 Mar 2004 &  &  &  &  &  &  & $6.2 \pm 1.9$ \\
01 Apr 2004 &  &  &  &  &  &  & $<8.3$ \\
05 Apr 2005 &  &  &  &  &  &  & $4.6 \pm 1.7$ \\
08 Apr 2005 &  &  &  &  &  &  & $8.0 \pm 3.2$ \\
13 May 2005 &  &  &  &  &  &  & $8.4 \pm 1.7$ \\
14 May 2005 &  &  &  &  &  &  & $10.4 \pm 3.8$ \\
15 May 2005 &  &  &  &  &  &  & $7.0 \pm 1.9$ \\
25 Mar 2006 &  &  &  &  &  &  & $<9.4$ \\
29 Mar 2006 &  &  &  &  &  &  & $<9.4$ \\
31 Mar 2006 &  &  &  &  &  &  & $<7.1$ \\
01 Apr 2006 &  &  &  &  &  &  & $<8.6$ \\
 \\
(Average) &  & $2.0 \pm 0.1$ & $2.2 \pm 0.1$ & $2.7 \pm 0.2$ & $2.9 \pm 0.4$ & $6.9 \pm 0.2$ & $6.9 \pm 0.6$\tablenotemark{c} &  & $0.21 \pm 0.04$ \\
($\chi^2/dof$) &  & 3.80  & 4.92  & 3.22  & 11.69  & 0.05  & 1.06\tablenotemark{c} &  & 25.81  \\
(Probability) &  & $<10^{-4}$ & $<10^{-4}$ & 0.002 & $<10^{-4}$ & 0.817 & 0.386\tablenotemark{c} &  & $<10^{-4}$ \\
\tableline
\end{tabular}
\end{center}

\tablenotetext{a}{Already published by \citet{Nagar:2001}.}
\tablenotetext{b}{Aperture size of 3\farcs8 (see Section~\ref{section:VLAspectra}).}  %
\tablenotetext{c}{Derived from only nine detected data sets.}
\tablenotetext{d}{Aperture size of 15\farcs1.}

\tablecomments{Column~1: observation date; Columns~2--9: flux densities at 1.7, 5, 8.4, 15, 22, 43, 100, and 347~GHz, respectively; Column~10:~spectral index, derived by the least-squares fit of $S_\nu \propto \nu^{+\alpha}$ at 5--22~GHz.  %
The last three lines: weighted average of flux densities at each frequency, reduced chi-square, and probability as statistics for the hypothesis of a constant value.}
}
\end{table*}

\subsection{NMA Observations}\label{section:NMAobservation}
We observed the nuclear regions at $\sim100$~GHz ($\lambda3$~mm) by using the NMA.  Array configurations were AB, C, and D with maximum baseline lengths of 351, 163, and 82~m, respectively.  
The visibility data were obtained with double-sided receiving systems.  
We use only the upper sideband at a center frequency of 100.777~GHz with a bandwidth of 1~GHz, which was practically line free; the emission lines such as HCN($J=$1--0) and HCO$^{+}$($J=$1--0) were in the lower sideband at a center frequency of 88.777~GHz.  For the data acquired in 2005 May, we use two sidebands centered at 89.729 and 101.729~GHz, both of which were practically line free.  
The typical system noise temperature was about 140~K in the double-sided band.  For gain calibration, we scanned between the target and a reference calibrator, 1150+497, every 20 or 25~minutes.  The flux scales of the calibrator were determined from relative comparisons with known flux calibrators such as Uranus.  In general, the uncertainty in the flux scale is expected to be 10\%--20\% as a result of accumulating the differences of residual errors between a reference calibrator and a flux calibrator.  These residual differences can be due to pointing error, seeing effect, calibrations of atmospheric attenuation and elevation-dependent antenna efficiency, and temporal variations in polarization and flux density of calibrators.  To avoid the residual errors of calibration from differential atmospheric attenuation and antenna efficiency, we observed 1150+497 and a flux calibrator at almost same elevations within 1~hr.  These observations were made within a few days of each other, typically, or were separated by up to 10~days in worst cases (from observations for NGC~4258); 1150+497 is a quasar with a large amplitude but gradual light curve on the timescale of years, which is much longer than the intervals between NGC~4258 and flux calibration observations.   In the present study, we assume a 15\% error in the flux scaling factor.  We observed bright quasars for about 30~minutes for bandpass calibration.  

The data were reduced using the {\tt UVPROC-II} package~\citep{Tsutsumi:1997} by standard means, including flagging bad data, baseline correction, opacity correction, bandpass calibration, and gain calibration.  
Each daily image was synthesized from visibilities in natural weighting and deconvolved using {\tt IMAGR} task of the Astronomical Image Processing System~({\tt AIPS}) developed at the NRAO.  

\subsection{VLA Data}\label{section:VLAdata}
In the VLA archive, we searched data sets of observations carried out in the continuum mode using multi-frequency bands with high spatial resolutions of A or B~configurations.  We retrieved (1)~all available data sets consisting of three or four bands in the range of 5--22~GHz ($\lambda\lambda1.3$--6~cm)\footnote{One of the data sets (AN8500) have already been published by \citet{Nagar:2001}.} for revealing spectral time evolution, (2) all available data sets including the 43~GHz~($\lambda7$~mm) band for the probe of millimeter spectrum, and (3) a data set including a deep imaging with A-array configuration at 1.7~GHz~($\lambda18$~cm) to evaluate the contamination from extended components of ``the anomalous arms'' in the photometric measurements at higher frequencies.  Observations of each data set were carried out at multi-frequency quasi-simultaneously within 3~hr typically, or longer (but within 1~day at most) hours for VLA data obtained as part of VLBI observations (BR0043, BG0062, BM0056, and BM0112).    

We reduced the data using the {\tt AIPS} by following the standard procedure.  3C~286 scans served as flux scaling factors.  We followed the guidelines for accurate flux density bootstrapping \citep{Perley:2003}, including correcting the dependence of the gain curve on elevation for the antennas and atmospheric opacity and by using the clean component models of 3C~286 in self-calibrations.  The bootstrap accuracy should be 1\%--2\% at 20, 6, and 3.6~cm, and perhaps 3\%--5\% at 2, 1.3, and 0.7~cm \citep{Perley:2003}.  For flux scaling factors, we assumed uncertainties of 2\% and 5\% at 1.7--8.4~GHz and 15--43~GHz, respectively.  All target images were synthesized with natural weighting and deconvolved with the {\tt AIPS IMAGR} task.

\subsection{JCMT Data}\label{section:JCMTdata}
We retrieved archival data for the 1998 March~18 observations of NGC~4258 that were acquired by the JCMT at 347~GHz ($\lambda$850~$\mu$m) in jiggle-map mode with the Submillimetre Common-User Bolometer Array \citep[SCUBA;][]{Holland:1999}.  
The data were processed with the SCUBA User Reduction Facility~(SURF) package.  We applied standard reduction procedures including flat fielding, flagging of transient spikes, correcting the extinction, pointing correction, sky removal, and flux density scaling.  A Uranus scan gave us a flux scaling factor, with an uncertainty assumed to be $\sim$15\% and an effective beam size of 15\farcs1.  On the blank sky, the measured root mean square~(rms) of the image noise was 10.5~mJy~beam$^{-1}$.

\section{Results}\label{section:results}
\begin{figure}
\epsscale{1.15}
\plotone{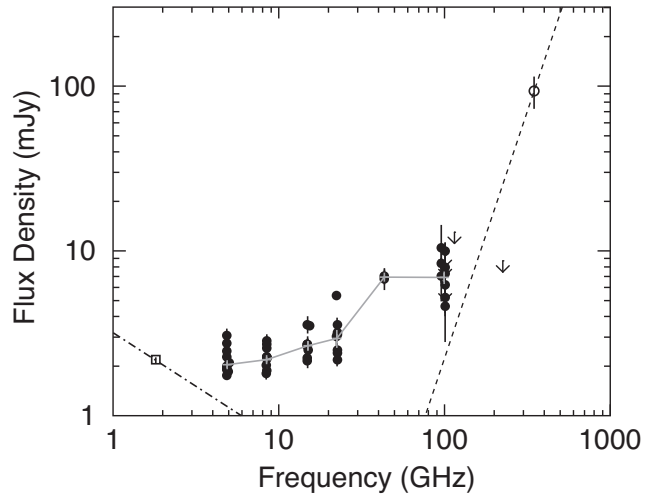}
\figcaption{Centimeter to submillimeter spectrum of the nuclear region of NGC~4258.  The data from VLA at 5--43~GHz (beam sizes of $\sim 0\farcs05$--3\farcs8) and NMA at $\sim100$~GHz ($\sim 3\arcsec$--10\arcsec) are represented as filled circles.  Gray crosses connected with gray solid lines represent weighted averages at each band.  The dashed curve represents a dust continuum spectrum (see Section~\ref{section:result_submm}) determined from the JCMT data at 347~GHz (15\farcs1).  The dot-dashed line represents an estimated upper limit of contribution from the extended emissions of the anomalous arms, which was derived from a flux measurement in a 3\farcs8~aperture at 1.7~GHz (open square; see Section~\ref{section:VLAspectra}).  The $3\sigma$~upper limits are represented by downward arrows.  Upper limits at 115 and 225~GHz are from literature (Section~\ref{section:nuclearmmemission}) \label{figure:N4258ALLplot}
}
\end{figure}

The results of the flux measurements are presented in Table~\ref{table:n4258result};   
Figure~\ref{figure:N4258ALLplot} shows the cm-to-submm spectrum of the NGC~4258 nucleus.    

\begin{figure}
\epsscale{1.15}
\plotone{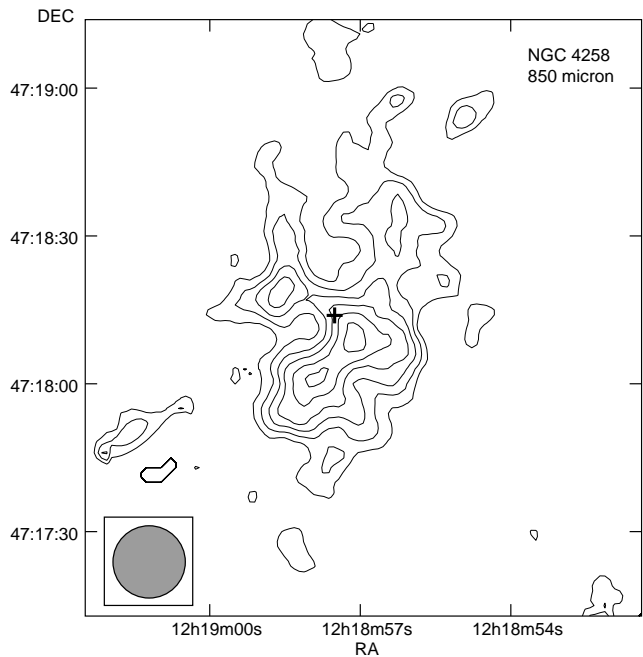}
\figcaption{JCMT image of the nuclear region ($\sim 4.3 \times 4.3$~kpc) of NGC~4258 at 347~GHz ($\lambda850$~$\mathrm{\mu}$m).  Contour levels are $\sigma \times$~(4, 5, 6, 7, 8, 9), where $\sigma =10.5$~mJy~beam$^{-1}$ is the rms image noise.  The cross symbol and its length represent the optical position of the NGC~4258 nucleus and pointing accuracy of the JCMT, respectively.  The beam size (15\farcs1) is shown in the lower-left corner of the image.\label{figure:NGC4258_850FINgray}}
\end{figure}

\subsection{Upper Limit of Dust Contamination}\label{section:result_submm}
The 347~GHz ($\lambda850\mu$m) JCMT image of the nuclear region with a field of view of $\sim2\arcmin$ shows a structure elongated along the major axis of the host galaxy with no apparent nuclear prominence (Figure~\ref{figure:NGC4258_850FINgray}).  Such a structure is very similar to that of carbon monoxide~(CO) line emission \citep{Regan:2001}.  These results suggest that most of the submillimeter emission does not originate in the AGN continuum but in interstellar dust associated with the host galaxy.  We measured the flux density from the region corresponding to the JCMT beam of $15\farcs1$~($\sim$530~pc) centered at the nucleus.  The continuum emission may be contaminated by the CO(3--2) emission of only $\sim1$~mJy in the SCUBA bandwidth of $\sim40$~GHz, according to the velocity-integrated intensity peak of the CO(1--0) map \citep{Helfer:2003} and the intensity ratio CO(3--2)/CO(1--0) \citep{Mao:2010}.

In general, the dust spectrum of external galaxies can be well represented via one- or two-temperature dust spectral models by using the modified blackbody emission $S_\nu \propto \nu^{\beta} B(T_\mathrm{D}$), where $B(T_\mathrm{D})$ is the blackbody spectrum at temperature $T_\mathrm{D}$, with $\beta \sim 1.6$--2 and two temperatures of $T_\mathrm{D}\sim20$ and $\sim43$~K; the lower-temperature component dominates at $<1000$~GHz in total flux \citep[][and references therein]{Temi:2004}.  
We adopt $\beta=1$ as an extreme case for conservative constraint at frequencies lower than 347~GHz of the JCMT data.  
As shown by a dashed line in Figure~\ref{figure:N4258ALLplot}, the dust contribution is expected to be $<2.4$~mJy at 100~GHz within a 15\farcs1 region, which means a strong upper limit for dust contamination toward the active nucleus of NGC~4258 in the flux measurements with smaller beam sizes of NMA ($<10$\arcsec) and VLA ($<3\farcs8$) at the lower frequencies \citep[cf.][]{Doi:2005,Doi:2011}.

\begin{figure}
\epsscale{1.15}
\plotone{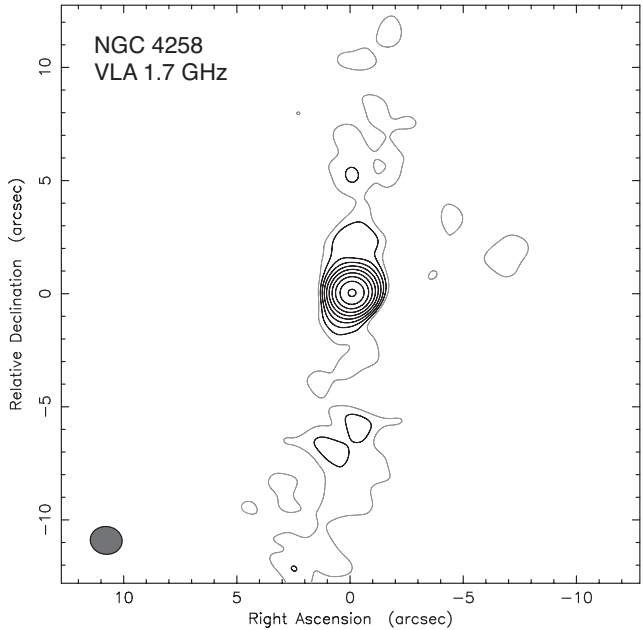}
\figcaption{VLA image with A-array configuration of the nuclear region of NGC~4258 at 1.7~GHz ($\lambda18$~cm) in natural weighting.  Contour levels in black curves are separated by factors of $\sqrt{2}$ beginning at $3\sigma$ of the rms noise $1\sigma=42\ \mu$Jy~beam$^{-1}$.  The gray curves represent $2\sigma$ contours.  The beam size of $1\farcs2 \times 1\farcs4$ at $\mathrm{P.A.}= 82\fdg6$ is shown in the lower-left corner of the image.\label{figure:NGC4258_1.4GHzVLAA}}
\end{figure}

\subsection{Nuclear Millimeter Emission}\label{section:nuclearmmemission}
NMA images at 100~GHz~($\lambda3$mm) with a field of view of $\sim 1 \arcmin$ revealed only a point-like single emission component at the nucleus.  We detected the emission for nine out of fifteen~epochs.  The flux densities were measured in the image domain by elliptical Gaussian profile fitting with the {\tt AIPS JMFIT} task.  We determined the total errors in the flux measurements from the root sum square of errors of the Gaussian fitting and flux scaling~(15\%; see Section~\ref{section:NMAobservation}).  For detected epochs, the millimeter continuum emission is significantly stronger than the centimeter emission.

The upper limit of dust contamination is $<2.4$~mJy at 100~GHz in a $\sim15\arcsec$ region at the nucleus~(Section~\ref{section:result_submm}).  
Moreover, an upper limit at 115~GHz was obtained on 1998 April~20 with a smaller beam size of $6\farcs1 \times 5\farcs4$ \citep{Helfer:2003}; an upper limit at 225~GHz was obtained on 2004 May~4 with an even smaller beam size of $3\farcs0 \times 2\farcs0$ \citep[][see Figure~\ref{figure:N4258ALLplot}]{Sawada-Satoh:2007}, indicating $<0.8$~mJy at 100~GHz (assuming $\beta\sim1$) as a conceivable compact dust component.  
On the other hand, the contamination from the low-frequency anomalous arms was estimated to be $<0.5$~mJy at 100~GHz on the basis of a spectral extrapolation using a flux density of 3.4~mJy at 4.9~GHz for the nuclear region of $14\arcsec$ \citep{Hummel:1989} and a spectral index of $-$0.65 of the anomalous arms \citep{Hyman:2001}.  Consequently, the contamination both from dust and anomalous arms can be practically neglected in our NMA measurements.  The detected millimeter emission is presumably dominated by the active nucleus.   

In a statistical sense, millimeter flux variability is unclear (cf.\ the last line in Table~\ref{table:n4258result}), which may be due to large uncertainty in flux scaling (see Section~\ref{section:NMAobservation}).

\begin{figure}
\epsscale{1.15}
\plotone{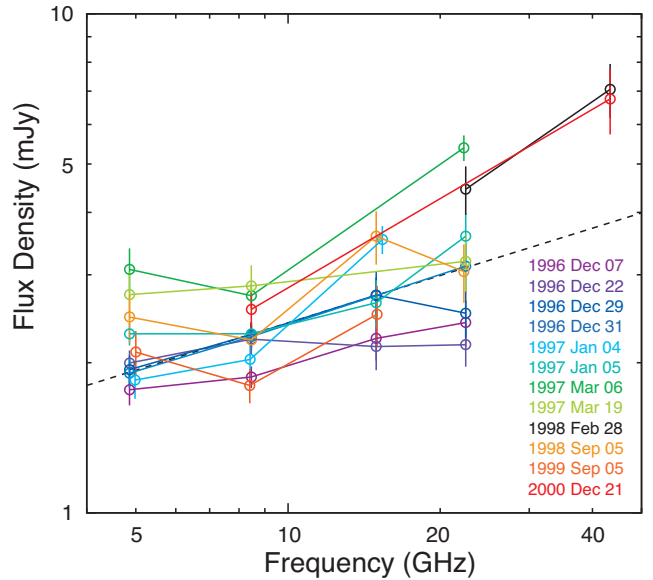}
\figcaption{Radio spectra for the NGC~4258 nucleus at 5--43~GHz obtained with the VLA A- and B-array configurations.  The dashed line represents the fit to a power-law spectrum ($S_\mathrm{1GHz}=1.2$~mJy, $\alpha=0.32 \pm 0.07$, where $S_\nu=S_\mathrm{1GHz} \nu^{+\alpha}$) for all data from 5 to 22~GHz.\label{figure:N4258_VLAmonitor}}
\end{figure}

\begin{figure}
\epsscale{1.15}
\plotone{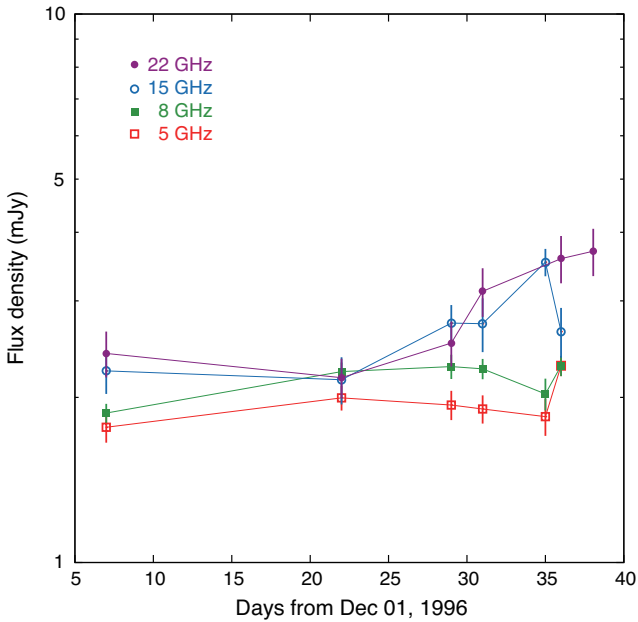}
\figcaption{Light curves of the NGC~4258 nucleus at 5--22~GHz in the first six epochs (and one more epoch at 22~GHz) with the VLA A-array configuration.  No 22-GHz measurement was obtained at the fifth epoch (the 35th day: 1997 January~4).\label{figure:N4258_VLAlightcurve}}
\end{figure}

\subsection{Spectral Evolution at VLA Wavelengths}\label{section:VLAspectra}
In VLA images at 5--43~GHz~($\lambda\lambda$6--0.7cm), we identified a single emission component at the nucleus in all cases analyzed.  No extended radio structure was apparent in these images with limited dynamic ranges.  The flux densities were measured with the {\tt AIPS JMFIT}.  The total errors were determined from the root sum square of errors of the Gaussian fitting and flux scaling (Section~\ref{section:VLAdata}).  In all cases, the emissions were almost unresolved; their deconvolved sizes were not beyond half of the FWHM of synthesized beams.  The deep VLA image at 1.7~GHz~($\lambda$18cm; Figure~\ref{figure:NGC4258_1.4GHzVLAA}) exhibits compact core + diffuse structures of the bases of ``anomalous arms'' to north--south directions.  This low-frequency image allowed us to evaluate the contamination from extended components with a steep spectrum into our photometry toward the nucleus at higher frequencies.  Photometry at 1.7~GHz with a 3\farcs8-aperture (larger than the largest beam size in VLA observations at 5--43~GHz; see Table~\ref{table:obslist}) resulted in $2.2 \pm 0.1$~mJy, which must include the emissions from both the nucleus and the bases of anomalous arms.  The two-point spectral index between 1.7--22~GHz (1997 January~7) is inverted ($\alpha = 0.20 \pm 0.11$), indicating a minor contribution from the anomalous arms ($\alpha=-0.65$; \citealt{Hyman:2001}) in the central region of 3\farcs8.  Hence, $2.2\times (\nu/1.7\ \mathrm{GHz})^{\alpha}$~mJy and $\alpha=-0.65$ presumably provides very conservative upper limits for estimating contaminations at higher frequencies as shown by a dot-dashed line in Figure~\ref{figure:N4258ALLplot}.

The VLA spectra for each epoch are shown in Figure~\ref{figure:N4258_VLAmonitor} and exhibit clear spectral variability with respect to a power-law spectrum~($\alpha=0.32 \pm 0.07$) represented by the dashed line, which is determined by the least-squares fit to all epochs in the range 5--22~GHz.  For each epoch, we also derived power-law spectral indices by least-squares fitting of quasi-simultaneously obtained data from 5 to 22~GHz~(Column~10~in Table~\ref{table:n4258result}), resulting in from $\alpha=0.06$ to $\alpha=0.64$; the average is $\langle \alpha \rangle = 0.21 \pm 0.04$.  The spectral index is not stable in a statistical sense (a probability $p<10^{-4}$ for the hypothesis of a constant value; see Table~\ref{table:n4258result}); however, it remains inverted (i.e., $\alpha > 0$) throughout all epochs.  Two-point spectra including 43~GHz ($\alpha=0.71\pm0.54$ between 22--43~GHz on 1998 February~28 and $\alpha=0.59\pm0.30$ between 8.4--43~GHz on 2000 December~21) produce evidence of this trend continuing up to millimeter wavelengths.  

The spectral features and their time evolutions were apparently complicated.  
Inverted spectra and, at the same time, flat and marginally steep spectra were locally observed in the frequency domain (Figure~\ref{figure:N4258_VLAmonitor}).  
The first six epochs show spectral features changing in only a month; the light curves for each frequency are shown in Figure~\ref{figure:N4258_VLAlightcurve}.  The flux variations of the two higher frequencies, 15 and 22~GHz, a gradual increase, appeared in the latter half of the six epochs.
On the other hand, flux variations of the lower two frequencies, 5 and 8.4~GHz, also correlated with each other but in a different way from those at the higher frequencies; a modest peak-out is apparent around the second or third epoch, and subsequently a (possible) sudden enhancement appears at the sixth epoch at both frequencies.  We estimate the size of the emitting region from $L < c \left| d\ln{S_\nu} / dt \right|^{-1}$, where $L$ is the size, $c$ is the speed of light, $S_\nu$ is the flux density, and $t$ is the observation time, assuming an exponential light curve \citep{Burbidge:1974,Valtaoja:1999}.  The synchronized increments by $\sim65$\% in light curves at 15 and 22~GHz during the last two weeks result in $L<$0.02~pc~($<$4000~AU or $<5300$Rs).  A flux decrease at 22~GHz over 13~days from 1997 March~6 to March~19 (beyond the time range of Figure~\ref{figure:N4258_VLAlightcurve}) results again in $L<0.02$~pc.  The flux density ranges and the timescales of variability are consistent with previous observations at 22~GHz (Appendix~\ref{section:previous_studies}).


\section{DISCUSSION}\label{section:discussion}
We found (1)~a complexly variable radio spectrum and (2)~the trend of a slightly inverted spectrum continuing throughout from centimeter (5~GHz, $\lambda6$cm) at least up to millimeter (100~GHz, $\lambda3$mm) regimes for the NGC~4258 nuclear radio continuum component.


\subsection{Radio Emitting Site}\label{section:radioemittingsite} 
We discuss the physical origin of observed radio emissions in our data.  
VLA beam sizes of $\sim0\farcs05$--3\farcs8, corresponding to 1.8--130~pc, decrease with increasing frequency; the observed inverted spectra are in opposite sense to the resolution effect.  It means the radio emission should be dominated by a compact component significantly smaller than the beam sizes.   
VLA and VLBI flux densities are almost the same \citep[e.g., $\sim3$~mJy at 22~GHz][]{Herrnstein:1998}, indicating the observed VLA flux densities originate in a sub-pc scale region.  
On the basis of VLBI images, the radio continuum flux comes from the northern side of the two-sided nuclear jet $\sim5$--10~mas away at 1.5~GHz \citep{Cecil:2000} and $\sim0.4$~mas (0.015~pc or 4000Rs) away at 22~GHz \citep{Herrnstein:1997} with respect to the dynamical center of the nearly edge-on viewed Keplerian disk of rotating water masers.  
Thus, most of the detected flux at 5--22~GHz in the present study is presumably of nuclear jet origin (rather than accretion flow), particularly from the northern side of jet, at locations $\ga4000$Rs away from the black hole.

Then, we consider the locations of 22--100~GHz radio emitting sites.  
Our finding of a slightly inverted spectrum is consistent with the picture of a relativistic plasma flow with an opacity gradient \citep{Blandford:1979}.  
The frequency dependence of the position of the ``core,'' which is identified by an intensity peak at the root of the jet \citep[i.e., the ``core-shift effect'';][]{Lobanov:1998}, is frequently observed for radio-loud AGNs \citep{Kovalev:2008,OSullivan:2009,Hada:2011,Sokolovsky:2011} and also for the LLAGN M81 \citep{Bietenholz:2004,Marti-Vidal:2011}.  
At a given frequency~$\nu$, the separation of the core from a central engine satisfies $r = \Omega \nu^k$ with $k \sim -1$ in most cases.  
The northern offset of $\sim0.4$~mas at 22~GHz in the NGC~4258 nuclear jet implies $\Omega\approx9$~mas~GHz ($\sim0.3$~pc~GHz or $\sim8\times10^4$Rs~GHz) with the assumption $k=-1$.  
According to this dependence, an expected position offset of $\sim6$~mas at 1.5~GHz is actually consistent with the observed offset of $\sim5$--10~mas at 1.5~GHz.  
Our finding of a continuous trend as an inverted spectrum at least up to 100~GHz suggests that the radial profile of jet structure maintains over the range of radio emitting sites of 1.5--100~GHz or higher 
because both the frequency dependence of core shift and spectral index are relevant to the radial profile of jet structure \citep{Blandford:1979}.   
The emitting site at 100~GHz is, therefore, expected to be $\sim800$Rs away, which is $\sim4.5$~times as close as the 22~GHz emitting site to the central engine; it is still far from an effective radius of $<100$Rs for an optically thick radio emitting photosphere of the ADAF \citep{Mahadevan:1997,Herrnstein:1998}.

Finally, we discuss emitting sites at even higher frequencies.  
We cannot determine a spectral profile at $>100$~GHz because the JCMT image at 347~GHz indicates heavy contamination by dust emission toward the nucleus (Section~\ref{section:result_submm}).   
On the other hand, infrared flux measurements by isolating the nucleus with high angular resolutions \citep{Chary:2000} are very suggestive.  
A flux density of 435~mJy at 17.9~$\mu$m~(16,800~GHz) is much higher than an extrapolation from the radio spectrum of $\alpha \approx 0.3$.    
It invokes a different origin, although the infrared--optical spectrum of $\alpha=-1.4\pm0.1$ indicates a nonthermal origin \citep{Chary:2000} as well.    
\citet{Yuan:2002a} proposed a jet-dominated SED model for the observed broadband SED that peaks in the infrared regime for NGC~4258.  The SED model ascribes the infrared peak and variable X-ray spectrum to synchrotron and self-Comptonized emissions, respectively, from the jet-base with a shock diameter of $5\mathrm{R_S}$, which is a preacceleration region in the jet connected with the innermost accretion flow.  Radio emissions cannot be attributed to the jet-base because of strong synchrotron self-absorption in a very compact region.  
An outer jet component as a postacceleration region in the jet on a scale of $>100\mathrm{R_S}$ \citep{Yuan:2002a} is required to explain the radio emissions.    
Using our finding, the VLBI results, and the SED model, two predictions can be made for future observations:   
(1)~the radio spectral variation will not be so promptly correlated with those in the infrared or X-ray regimes \citep[cf.][for M81]{Miller:2010} because of significantly different linear scales of emitting sites and (2)~the outer jet component will provide a contribution comparable with that of the jet base at submillimeter wavelengths, which implies a spectral upswing from $\alpha \approx 0.3$ at lower frequencies to $\alpha \la 2.5$ at higher frequencies rather than a spectral turnover.

 
\subsection{Mechanism of Radio Spectral Variation}\label{section:mechanism}
The complexly variable inverted spectrum we found should reflect the nature of the northern side of the nuclear jet, where a large fraction of radio emission arises (Section~\ref{section:radioemittingsite}).    
The time variation of bright systemic masers is strongly correlated with the northern jet emission; 
it is naturally explained by intrinsic variability in the background continuum of the southern jet and only mildly saturated masers \citep{Herrnstein:1997}.  
It means that the observed radio properties in the northern jet mainly originate in intrinsic ones, rather than in the external effect such as free--free absorption.  
On the geometry of a warped disk model \citep[see fig.~10 in][]{Herrnstein:2005}, the northern jet is behind the disk at radii of $>0.29$~pc in line of site, where intervening medium is not ascribable to an effective free--free absorber in terms of the opacity \citep{Neufeld:1995} or timescale \citep[an inhomogeneity of $\sim10^{15}$~cm in size and a rotating velocity of $<750$~km~s$^{-1}$;][]{Fruscione:2005}.

In this framework, the frequency dependence of the core shift in the northern jet (Section~\ref{section:radioemittingsite}) is ascribable to synchrotron self-absorption; the emitting site at 22~GHz would be almost opaque at 5~GHz.  
The light curves at lower two frequencies were very similar; those at higher two frequencies formed in synchronization.  However, the former pair showed a different trend from the latter pair (see Figure~\ref{figure:N4258_VLAlightcurve} in Section~\ref{section:VLAspectra}).  
This behavior is consistent with a picture of different emitting regions at different frequencies.      
The observed complexly-variable inverted spectrum in the NGC~4258 nucleus can be interpreted as the superposition of emission spectra originating at different locations in the nuclear jet with frequency-dependent opacity.  
Even if a phenomenon propagates from upstream to downstream at the speed of light, a variation at 15 and 22~GHz would appear as a corresponding variation at 5 and 8.4~GHz a few months later.   
Such behavior was not detectable in our data for the first month (Figure~\ref{figure:N4258_VLAlightcurve}); a series of the later epochs (1997 March onward) were too sparse.  

\subsection{Comparisons with M81}\label{section:comparison}
Our findings of the radio spectral properties for the NGC~4258 nucleus allow us to understand that the underlying mechanism in the jet component (Sections~\ref{section:radioemittingsite} and \ref{section:mechanism}) is basically identical to that of M81.    
This is because the two LLAGNs share all the following observational aspects in radio regimes: 
(1)~a variable and complex spectral profile, 
(2)~slightly inverted spectrum with $\alpha \approx 0.3$ in average, and 
(3)~core shift on a nuclear jet (see Section~\ref{section:introduction} for M81).       
Presumably, the underlying mechanism must be compatible to other more distant LLAGNs (spatially unresolved so far) too, whose nuclear components are rapidly varying radio emissions \citep{Anderson:2005} showing flat/slightly inverted spectra \citep{Nagar:2001,Doi:2005}.

\begin{figure}
\epsscale{1.15}
\plotone{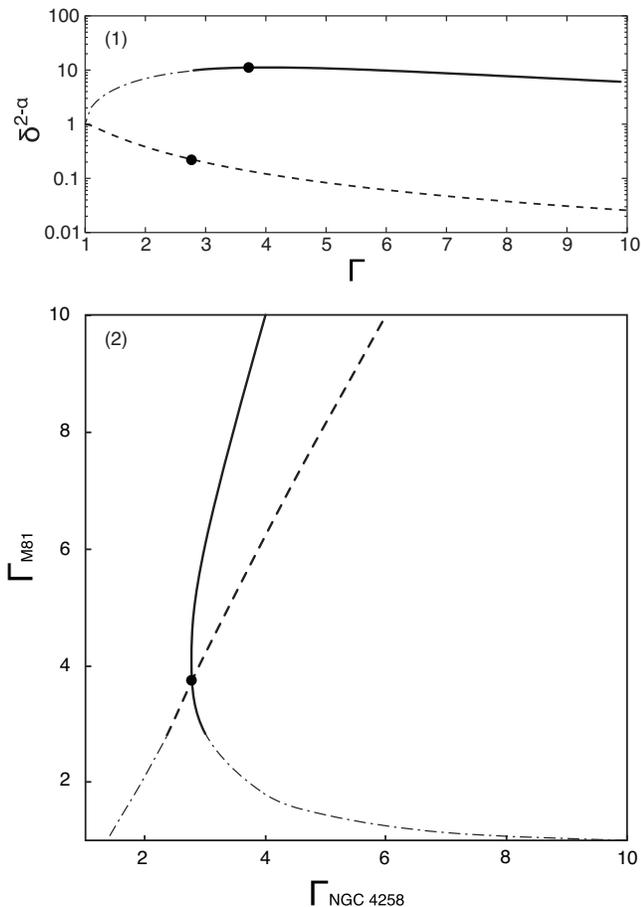}
\figcaption{(1)~Boosting factors of Doppler beaming in flux densities for continuous jets.  Solid and dashed curves represent the cases of inclinations of 14\degr and 82\degr, respectively (Equation~(\ref{eq:jet-couter_ratio})).  Filled circles represent possible solutions for M81 and NGC~4258 (Section~\ref{section:comparison}).  (2)~Relationship between bulk Lorentz factors of M81 and NGC~4258.  The solid curve represents Equation~(\ref{eq:delta}), a constraint from the ratio of radio loudness of M81 and NGC~4258.  The dashed line represents Equation~(\ref{eq:omega}), a constraint from the ratio of observed core shifts of M81 and NGC~4258.  The range ($\Gamma_\mathrm{(M81)}<2.8$) illustrated by dot-dashed curves would not be allowed because of the observed one-sidedness.\label{figure:gamma_solution}}
\end{figure}

Now, the two well-studied LLAGNs enable us to supply a physical understanding by means of quantitative comparisons.  
NGC~4258 exhibits aspects that differ from M81 in terms of:   
(1)~the apparent flux ratio of jet to counter jet $R_\mathrm{F}$: two-sided morphology in NGC~4258 ($R_\mathrm{F} \approx 6$; \citealt{Herrnstein:1998}), whereas one-sided in M81 ($R_\mathrm{F} >170$), 
(2)~radio loudness (radio-to-X-ray luminosity ratio) $R_\mathrm{X} \equiv \nu P_\mathrm{radio}/L_\mathrm{X}$ \citep{Terashima:2003}: a much radio-louder SED for M81 ($R_\mathrm{X} \approx 2 \times 10^{-3}$) compared to NGC~4258 ($R_\mathrm{X} \approx 4 \times 10^{-5}$), and 
(3)~magnitude of core shift: $\Omega\approx9$~mas~GHz ($\sim0.3$~pc~GHz or $\sim8\times10^4$Rs~GHz) for NGC~4258 (\citealt{Herrnstein:1997,Cecil:2000}; see Section~\ref{section:radioemittingsite}) is much larger than $\Omega=1.75\pm0.20$~mas~GHz (0.031~pc~GHz or 4500Rs~GHz) for M81 \citep{Marti-Vidal:2011}.      
To calculate $R_\mathrm{F}$ of M81, we adopted a peak intensity of $\sim100$~mJy~beam$^{-1}$ and three times the image noise of $\sigma\sim0.2$~mJy~beam$^{-1}$ in the images at 8.4~GHz by \citet{Bietenholz:2000}.  
To calculate $R_\mathrm{X}$, we used observed flux densities at 22~GHz of typically $\sim3$~mJy and $\sim100$~mJy for NGC~4258 (Table~\ref{table:n4258result}) and M81 \citep[e.g.,][]{Bietenholz:2000}, respectively, as representative of radio data because of similar spectral indices ($\alpha \approx 0.3$) and 2--10~keV X-ray luminosities typically $\sim1\times10^{41}$~erg~s$^{-1}$ and $\sim2\times10^{40}$~erg~s$^{-1}$ for NGC~4258 \citep{Makishima:1994,Fruscione:2005,Yamada:2009} and M81 \citep{Ishisaki:1996,Markoff:2008,Miller:2010}, respectively.  
In the present paper, we demonstrate that the differences in (1)--(3) can be attributed to inclination effects by following discussions; 
M81 is presumably a pole-on viewed system at an inclination of $\theta=14\degr$~\citep{Devereux:2003}, whereas NGC~4258 at $\theta=82\degr$ is viewed nearly edge-on \citep{Herrnstein:1996}.

First, we examine the apparent flux ratio of jet to counter jet $R_\mathrm{F}$.  
The Doppler beaming effect would make emission boosted by a factor of $\delta^{2-\alpha}$ in a continuous flow \citep{Ghisellini:1993}, where the Doppler factor and bulk Lorentz factor are defined as $\delta \equiv [\Gamma (1-\beta\cos{\theta})]^{-1}$ and $\Gamma \equiv (1-\beta^2)^{-0.5}$, respectively, and $\beta$ is the jet speed in the unit of speed of light.  
The apparent flux ratio of jet to counter jet is expressed as 
\begin{equation}
R_\mathrm{F} = \left( \frac{1+\beta \cos{\theta}}{1-\beta \cos{\theta}} \right)^{2-\alpha}.\label{eq:jet-couter_ratio}  
\end{equation}
With this equation, $\theta=82\degr$ for NGC~4258 implies $R_\mathrm{F}<1.61$ at any value of $\Gamma$ (i.e., almost symmetric); the observed value of $R_\mathrm{F} \approx 6$ must be affected by free--free absorption on the southern jet \citep{Herrnstein:1996}.  
On the other hand, M81 with $R_\mathrm{F}>170$ requires $\Gamma>2.8$ at $\theta=14\degr$, and then, $\delta>3.8$.  
Thus, we can understand the two-sidedness in NGC~4258 as an inevitable result, and obtain a constraint of jet speed for M81.

Second, we verify the difference in radio loudness $R_\mathrm{X}$.  
At these inclinations, the Doppler beaming effect makes the (approaching) jet emission {\it boosted for M81 and deboosted for NGC~4258} {\it at any bulk Lorentz factors} (Figure~\ref{figure:gamma_solution}, panel~1).  
Therefore, we can naturally understand that M81 is significantly radio louder in SED than NGC~4258, if radio and X-ray emissions originate in a beamed and a non-beamed component, respectively.  
Indeed, studies of SED modeling ascribed radio and X-ray emissions to the outer jet (synchrotron) as a postacceleration region and the jet-base component (inverse Compton) as a preacceleration region, respectively, in both cases of NGC~4258 \citep{Yuan:2002a} and M81 \citep{Markoff:2008}.  
As a result, for NGC~4258, the jet-base component dominates throughout from IR and X-ray \citep{Yuan:2002a}, which is presumably due to deboosting on the outer jet.  
On the contrary, the outer jet plays a major role in the SED of M81 \citep{Markoff:2008}.  
Here we attempt to constrain the bulk Lorentz factors of the nuclear jets (corresponding to outer jets in the SED models).  
We consider an observed radio power as a boosted emission ($P_\nu^\mathrm{obs} = \delta^{2-\alpha} P_\nu^\mathrm{int}$) at a given frequency $\nu$.  
We assume that the intrinsic radio power ($P_\nu^\mathrm{int}$) is relevant to a total radiated synchrotron luminosity ($L_\mathrm{syn}$).  Moreover, we adopt a tentative assumption of $L_\mathrm{syn} \propto L_\mathrm{jet} \propto L_\mathrm{X}$, where $L_\mathrm{jet}$ and $L_\mathrm{X}$ are the total jet luminosity and X-ray luminosity, respectively, as non-beamed ($\Gamma \sim 1$) component.    
Then, 
\begin{eqnarray}
\frac{R_\mathrm{X(NGC4258)}}{R_\mathrm{X(M81)}}
&=&  
\displaystyle \frac{ \frac{P^\mathrm{obs}_{\nu\mathrm{(NGC4258)}}}{L_\mathrm{X (NGC4258)}} }{ \frac{P^\mathrm{obs}_{\nu\mathrm{(M81)}}}{L_\mathrm{X (M81)}} } \nonumber \\
&\approx& 
\left( \frac{\delta_\mathrm{(NGC4258)}}{\delta_\mathrm{(M81)}} \right)^{2-\alpha}.\label{eq:delta}  
\end{eqnarray}
Under fixed inclinations $\theta$, Lorentz factors $\Gamma_\mathrm{(M81)}$ and $\Gamma_\mathrm{(NGC4258)}$ in Doppler factors $\delta$ can constrain each other as represented by a solid curve in Figure~\ref{figure:gamma_solution}, panel~2.  
We can find $\Gamma_\mathrm{(NGC4258)} \ga 3$ based on Equation~(\ref{eq:delta}) together with the constraint from the observed one-sidedness in the M81 nuclear jet (the dot-dashed curve in Figure~\ref{figure:gamma_solution}, panel~2 is in the forbidden region).

Third, we discuss the difference in magnitude of core shift $\Omega$.  
The value of $\Omega$ also depends on both inclination angle $\theta$ and Lorentz factor $\Gamma$.  We employ Equation~(11) in \citet{Lobanov:1998} and then consider the ratio of $\Omega$ for NGC~4258 and M81: 
\begin{eqnarray}
\Omega \propto \frac{(1+z) \sin^{5/3}\theta}{D_\mathrm{L}\phi} \left( \frac{L_\mathrm{syn} \delta}{\Gamma \sqrt{\Gamma^2-1} \Theta} \right)^{2/3}, \nonumber \\ 
\frac{\Omega_\mathrm{(NGC4258)}}{\Omega_\mathrm{(M81)}} = \frac{9\ \mathrm{mas\ GHz}}{1.75\ \mathrm{mas\ GHz}},\label{eq:omega}       
\end{eqnarray}
where $D_\mathrm{L}$ is the luminosity distance.  
We assume that $\phi$ and $\Theta$, the opening angle of jet and the range of emission region along jet ($\Theta \equiv \ln{(r_\mathrm{max}/r_\mathrm{min})}$; \citealt{Blandford:1979}), respectively, are the same between NGC~4258 and M81, and then eliminated in the ratio in Equation~(\ref{eq:omega}).  We assume $L_\mathrm{syn} \propto L_\mathrm{X}$ again.  
Under fixed inclinations $\theta$, Lorentz factors $\Gamma_\mathrm{(M81)}$ and $\Gamma_\mathrm{(NGC4258)}$ involved in Equation~(\ref{eq:omega}) can constrain each other as represented by a dashed line in Figure~\ref{figure:gamma_solution}, panel~2.  
The dependence suggests that the jet speed is not so different but may be a slightly lower speed in NGC~4258.

Thus, the combination of these three ratios in terms of (1)~$R_\mathrm{F}$, (2)~$R_\mathrm{X}$, and (3)~$\Omega$ constrains Lorentz factors:      
the intersection of the two equations~(Equations~(\ref{eq:delta}) and (\ref{eq:omega})) is at $\Gamma_\mathrm{(NGC4258)}\sim2.8$ ($\beta\sim0.932$ and $\delta\sim0.42$) and $\Gamma_\mathrm{(M81)}\sim3.7$ ($\beta\sim0.963$ and $\delta\sim4.1$), shown as a filled circle in Figure~\ref{figure:gamma_solution}, panel~2 and corresponding values of $\delta^{2-\alpha}$ in Figure~\ref{figure:gamma_solution}, panel~1.  
This $\Gamma_\mathrm{(M81)}$ value is consistent with $\Gamma>2.8$ derived from the observed one-sidedness using Equation~(\ref{eq:jet-couter_ratio}).  
Thus, {\it jet speeds in both M81 and NGC~4258 are fairly relativistic},  
although these discussions are based on several assumptions.  
Roughly speaking, a lower ratio in Equation~(\ref{eq:delta}) leads to a larger $\Gamma_\mathrm{(M81)}$; a smaller ratio in Equation~(\ref{eq:omega}) leads to a larger $\Gamma_\mathrm{(NGC4258)}$; $\Gamma_\mathrm{(M81)}>\Gamma_\mathrm{(NGC4258)}$ is always seen in a reasonable range of $\Gamma \approx 1.5$--30 and a conceivable uncertainty of $\sim\pm50$\% in the ratios.  If we adopt another assumption as $L_\mathrm{syn} \propto L_\mathrm{2-10keV}^{0.69} M_\mathrm{BH}^{0.61}$ \citep[the revised relation of the fundamental plane by][]{Plotkin:2012} instead of $L_\mathrm{syn} \propto L_\mathrm{2-10keV}$, $\Gamma_\mathrm{(NGC4258)}\sim3.2$ and $\Gamma_\mathrm{(M81)}\sim6.8$ are obtained.

Consequently, the observed differences between NGC~4258 and M81 in (1)--(3) can be naturally interpreted by the Doppler beaming effect due to fairly relativistic jet speeds and quite different inclinations.  
In previous studies, $\Gamma\ga2$--3 was also predicted for NGC~4258 \citep{Falcke:1999,Yuan:2002a} and M81 \citep{Falcke:1996,Markoff:2008} by the model of the outer jet freely expanding and weakly accelerated by internal pressure gradients, which can naturally explain the frequency-dependent size and spectral index of compact radio core \citep{Falcke:1995}.  %
In fact, mildly or fairly relativistic jet speeds have been inferred from observations from other LLAGNs/nearby Seyfert galaxies, e.g., Arp~102B \citep[$\geq 0.45c$;][]{Fathi:2011}, NGC~4278 \citep[$\sim0.76c$ ($\Gamma\approx1.5$);][]{Giroletti:2005}, and NGC~7674 \citep[$\sim0.92c$ in projection;][]{Middelberg:2004}.


\section{SUMMARY}\label{section:summary}
We investigated the spectral properties of the second closest LLAGN radio nucleus in NGC~4258 and discussed the physical origin of these properties and comparisons with the closest LLAGN M81.  The research is summarized by the following points: 
\begin{itemize}
\item Radio flux variability was detected at all bands in the centimeter regime (5--22~GHz).  In a statistical sense, millimetric variability is unclear, which may be due to an insufficient number of epochs (43~GHz) and the large uncertainty in flux measurements (100~GHz).  The VLA 1.7~GHz deep image and the JCMT 347~GHz observation provided the upper limits of extended synchrotron and dust contaminations, respectively.  
\item Time-averaged flux densities showed a slightly inverted spectrum ($\alpha \sim 0.3$) at 5--100~GHz.  A spectral turnover at a higher frequency is unclear from the available data. 
\item Quasi-simultaneous multi frequency VLA observations at 5, 8, 15, and 22~GHz revealed a variable and complicated spectral profile, but always an inverted spectrum in all 10 epochs.  The first six epochs contained intra-month variation.  The light curves at lower two frequencies were very similar; those at higher two frequencies formed in synchronization.  However, the former pair showed a different trend from the latter pair.  
\item The observed radio spectral properties can be interpreted as the superposition of emission spectra originating at different locations with frequency-dependent opacity along the nuclear jet.%
\item The NGC~4258 nucleus demonstrates the following characteristics similar to those of M81: (1)~a complexly variable spectral feature, (2)~a slightly inverted spectrum with $\alpha \approx 0.3$ in average, and (3)~apparent core shift in the AU-scale jet.  %
\item We made quantitative comparisons between NGC~4258 and M81 in terms of jet/counter jet ratio, radio loudness, and magnitude of core shift.  We found a solution of the combination of bulk Lorentz factors $\Gamma_\mathrm{(NGC4258)}\sim2.8$ and $\Gamma_\mathrm{(M81)}\sim3.7$ for inclinations of 82\degr\ and 14\degr, respectively.  These discussions are based on several assumptions; however, these results are consistent with relativistic jet speeds inferred in a few other LLAGNs in literature.  %
\end{itemize}

\acknowledgments
We are grateful to the anonymous referee for offering constructive comments that have contributed in substantially improving this paper.  We acknowledge the support of the NRO staff for operating the telescopes and for their continuous efforts to improve the performance of the instruments.  We especially acknowledge S.~K.~Okumura and Y.~Tamura for helpful advice in observations with the NMA.  NRO is a branch of the National Astronomical Observatory of Japan (NAOJ), which belongs to the National Institutes of Natural Sciences (NINS).  AD thanks M.~Kino for useful discussions.  We have made extensive use of the VLA archival data retrieved from the NRAO, which is a facility of the National Science Foundation operated under cooperative agreement by Associated Universities, Inc.  In addition, we have made use of the JCMT archival data retrieved from the Canadian Astronomy Data Centre, which is operated by the Dominion Astrophysical Observatory for the National Research Council of Canada's Herzberg Institute of Astrophysics.  This work was partially supported by a Grant-in-Aid for Young Scientists~(B; 18740107, A.D.) and Grant-in-Aid for Scientific Researches~(C; 21540250, A.D.~and B; 24540240, A.D.) from the Japanese Ministry of Education, Culture, Sports, Science, and Technology~(MEXT).  This work was partially supported by the Center for the Promotion of Integrated Sciences~(CPIS) of Sokendai.

%

\appendix
\section{Previous Radio Continuum Observations}\label{section:previous_studies}

{\it Spectrum (by quasi-simultaneous observations)}.  \citet{Ho:2001} reported a slightly resolved core of $S_\mathrm{1.4GHz}=2.73$ and $S_\mathrm{5GHz}=2.18$~(mJy), $\alpha=-0.18$, in matched angular resolutions of $\sim1\arcsec$, which were in agreement with previous observations at a similar resolution \citep{Vila:1990}.
\citet{Hyman:2001} reported the time variation of spectral indices ranging from $-0.18$ to $-0.41$ between 1.4--5~GHz toward the nucleus using the VLA A-array configuration.  On the other hand, the spectral index map shows a relatively uniform distribution of the steep spectrum of $\alpha=-0.65 \pm 0.10$ throughout ``the anomalous arms''.  
An inverted spectrum had been seen toward the nucleus at higher frequencies: $S_\mathrm{5GHz}=2.4$ and $S_\mathrm{15GHz}=3.2$~(mJy), $\alpha=+0.4$, in matched resolutions of $\sim3\arcsec$ \citep{Turner:1994}.  
\citet{Nagar:2001} reported a folded spectral feature (2.1, 1.9, and 2.8~mJy at 5, 8.4, and 15~GHz, respectively) by the VLA A-array observations at matched resolutions of $0\farcs5$.  

{\it Variability}.  Significant flux variability has been previously suggested at 22~GHz: the continuum component in VLBI images seems to vary by a factor of several times ($S_\mathrm{22GHz}=0.7$--4.0~mJy) during 15~epochs in the period 1997.18--2001.61, although these measurements are based on a limited range of baseline lengths \citep{Argon:2007}.  \citet{Herrnstein:1997} also showed flux variability of the continuum component at 22~GHz from $<2$ to 6~mJy based on both VLA and VLBI in the period 1994.3--1996.3.  The emission varied by up to 100\% over the timescale of a few weeks.

\bibliography{N4258_ApJ_20130128_astrp-ph2.bbl}

\end{document}